\documentstyle[aps,version2,preprint]{revtex}

\begin{document}

\begin{title}

Correlations between Aharonov-Bohm effects and one-dimensional\\
subband populations in GaAs/Al$_{x}$Ga$_{1-x}$As rings
\end{title}

\author{J.\ Liu$^1$, W.X. Gao$^1$, K.\ Ismail$^{2,3}$, K.Y.\
Lee$^{3}$, J.M.\ Hong$^3$ \\
and S.\ Washburn$^1$}

\vspace{12pt}
\begin{instit}
$^1$ Dept. of Physics and Astronomy,
The University of North Carolina at Chapel Hill,\\
Chapel Hill, NC 27599-3255, USA\\
$^2$ Dept.\ of Telecommunication and Elec.\ Eng., University of Cairo,\\
Cairo, Egypt.\\
$^3$ IBM Watson Research Center, Yorktown, NY 10598, USA.
\end{instit}



\begin{abstract}
\vspace{-6pt}

The Aharonov-Bohm (AB) interference patterns in ring-shaped conductors
are usually dominated by random features.  The amplitude of the oscillations
is random from sample to sample and from point to point on the magnetic field
axis owing to random scattering of the electron trajectories by impurities
within the wires.  We report experiments on new devices made with wet etching
and global gates, which have shown major progress towards removing the random
features.  In loops that exhibit ballistic conductance plateaux and cyclotron
orbit trapping at $4.2K$, the random pattern of AB oscillations (observed
for $T < 0.1K$) can be replaced by much more ordered one -- especially if only
a few transverse modes are populated in the ring.  The amplitude and shape of
the oscillation envelope function change systematically as subbands are
populated in the wires forming the loops.  Mechanisms governing the AB effect
in the ballistic regime are discussed. Correlation has been found between
the $G(V_{g},B=0)$ staircase and the ``beating period" of the envelope
functions.  Quantum oscillations in $G(V_{g},B = 0)$ are consistent with
direct interference of paths of unequal length.  Both the correlations and the
quantum oscillations in gate voltage are signatures of ballistic transport.

\vspace{24pt}

\today,    Submitted to Physical Review B

\end{abstract}


\begin{center}
{\bf \center Introduction}
\end{center}
\vspace{10pt}

The Aharonov-Bohm (AB) interference effects\cite{AB}
have been studied extensively in small metallic
rings.\cite{WW}   In these devices, electrons encounter large amounts of
elastic
scattering and move diffusively.  The elastic mean free path is typically
$l_e \sim 10-100$\AA.  The AB effects are observable in these systems
because at low temperatures the
quantum phase information is retained over a much longer distance $l_\phi$,
the phase coherence length,\cite{kltheory,Imry} which can be 3 to 4 orders of
magnitude greater than $l_e$ at temperatures below $T=1K$.
In metals the electrons are highly degenerate
and the Fermi wave length is much smaller than the diameters of
wires that form the loop (typically 3 orders of magnitude),
hence there are a large number of states on Fermi surface contributing
to the electron transport, and there is no quantization of electron motion
in the direction transverse to the wire.

Following the discovery of the quantized conductance in
semiconductor heterostructure point contacts\cite{Imry,BvW},
much work has been devoted to the study of one-dimensional (1-d) electron
subbands (or modes) in electron transport.\cite{Timpqpc,bh}
The subbands arise due to the low carrier density in
semiconductor interfaces, which leads to a long Fermi wavelength, which is
in turn comparable to the width $t$ of the wire.
Both theory and experiments\cite{Imry,Timpqpc,bh,GKrev} have shown that, in
a straight channel, the conductance contributed  by
each subband is $2e^{2}/h$ (2 from spin degeneracy).
If the carrier density is controlled by a gate voltage,
the conductance will change in increments of the basic conductance
quanta $2Ne^{2}/h$, where $N = 0,1,2...$ is the number of occupied subbands.
Recently, AB experiments have been carried out in loops fabricated on
high electron mobility GaAs/Al$_{x}$Ga$_{1-x}$As
heterostructures.  In this type of device, large amplitude
AB oscillations have been
reported by several groups.\cite{Timp,GAloops,ford88,TimpQHE,PGNdV,CJBFk}
Since electron conduction is via the subbands, the subband population
ought to affect the AB interference pattern dramatically.
If no scattering occurs, and electrons are guided only by
the electrostatic confinement that defines the shape of the device and
the Lorentz force, we would expect the electron
transmission to be very similar to microwave transmission through
waveguides, and that changing the
electron density would be analogous to tuning the microwave frequency.
In a pure ballistic situation, such a device would be a true
{\em solid state interferometer} in a magnetic field.

Consider what happens to the interference
when the mode population changes.
At zero magnetic field, conductance $G(V_{g},B=0)$ is a
mode-counting staircase in $V_g$. In the plateaux (the tread of the
``staircase''), the modes are well
defined (transverse momentum is the conserved quantum number),
and in the transition regions or the risers, new modes are just turning
on.
At low enough fields, we may suppose that the
magnetic field does not alter the subband population substantially.
Correlations should be expected between AB oscillations at fixed gate
voltage $G(V_g =const,B)$ and the mode-counting staircase $G(V_{g},B=0)$:
for $V_g$ on the plateaux of the mode-counting staircase,
the AB oscillations
are larger and more ordered than $V_g$ in the risers.
As a result, the large ordered and somewhat small and disordered patterns
will appear alternatingly in the AB oscillations as we sweep the gate voltage.
At the same time, we also expect that as more and more modes are populated
the structure of AB effect will become more and more complicated.
This can
be understood in analogy with a waveguide operated in a single-mode
(or few-mode) transmission and in multi-mode transmission.
Earlier experiments\cite{CJBFk}, however, did not find the expected
correlations. Apparently the interference pattern was dominated by random
scattering from donors and surface defects,\cite{bh,Timpqpc}
and our ballistic picture simply does not apply.
(An exception occurred at {\em high} magnetic fields in the
regime of the quantized Hall effect, where beautiful regular oscillations
were observed from a single point contact
in the transition between two successive Hall plateaux\cite{bh}).

This paper will report the results from devices in which
scattering has been eliminated to a large degree.
Correlations have been found between AB interference patterns
and subband population.
In addition, quantum oscillations were seen
in $G(V_{g})$ at $B = 0$, which are consistent with direct interference
of paths of unequal length.
These are signatures of pure ballistic transport, and so our results contain an
encouraging step toward the ultimate goal of completely ballistic devices.

\vspace{10pt}
\begin{center}
{\bf Experiment }
\end{center}
\vspace{10pt}

Our samples were fabricated on a standard high-mobility
GaAs/Al$_{x}$Ga$_{1-x}$As modulation-doped layer ($x = 0.3$,
carrier density $n_s = 2 \times 10^{15}/m^2$ and
mobility $\mu = 90 m^2/Vsec$) grown by molecular beam epitaxy.  The ring
geometries was defined by shallow mesas formed through a wet etching
technique.  Metal gates cover all active portions of the devices.
We attribute the excellent quality of these devices directly to the softness
of the etching and screening effect of the global gate.  The effect of the
gate in reducing the long range interaction of ionized
dopants should not be underestimated.
More details of the fabrication are reported elsewhere.\cite{washburn91}
An earlier paper by the authors discussed an experiment
on a similar single ring device 4.2K. Although that temperature was too
high to see any AB oscillations, cyclotron orbit trapping was observed, which
in turn is convincing evidence of substantial ballistic transport
in the device.\cite{liu93}

The samples we used are single ring and coupled two rings and four rings.
The lithographic geometry for the coupled two rings is illustrated in
of Figure 1.
The two parallel rings are ``joined at the hip''.
The four ring sample are the obvious extension of the series with four
parallel rings.  The rings in all the sample are of the same size.
The lithographic pattern has the center radius
of $r = 0.80 \mu m$ and line width of $t =
0.4 \mu m$.
Previous analysis of the data in similar samples leads to a more realistic
estimate for width of the conduction channel to be
$t = 0.3 \mu m$.\cite{liu93}.
The temperature at which the data was taken was $0.04K$
unless specified.  This temperature is in the regime of large phase coherence
length, where interference effects are quite apparent.
The resistance was measured through standard low frequency ac technique
with PAR 124A Lock-in amplifiers and Ithaco 1211 or 1212 current amplifiers.
The excitation voltage ranged from $1 \mu V$ to $4 \mu V$ at the
frequency of $13$ or $278.5$Hz. During the measurements, the samples were
immersed in the mixing chamber of the dilution refrigerator to ensure
good thermal contact.

\vspace{10pt}
\begin{center}
{\bf Analysis of the magnetoresistance}
\end{center}
\vspace{10pt}

In Figure 2 the lower curve displays a magnetoresistance data for the
single ring in a ``random'' state.
At this $V_g$ there are $5.5$ subbands are filled
(judged by the conductance $G(V_g,B=0)$, which is not shown)
in the ring.  The interference pattern in this state
is similar to data from earlier experiments in
heterojunctions\cite{Timp,GAloops,ford88,CJBFk,TimpQHE,PGNdV},
and also from diffusive metallic loops\cite{WW}.
The phase and amplitude of the oscillations are uncorrelated
from point to point along the magnetic field axis.
The $h/je$ (j is an integer)
frequencies are mixed with considerable amounts of other ``frequencies"
of oscillation yielding a random amplitude for the AB
oscillations.  In the ``random'' state, no correlations among
the data at different $V_g$ or between the AB oscillations and
the $G(V_g,B=0)$ staircase are apparent.

In Figure 2 the top curve displays a typical magnetoresistance
data for the single ring sample in the ``ordered'' state.
$V_g = 0.15V$, which corresponds to having 2
subbands filled in the ring.
In contrast, it is dramatically different from the bottom curve,
and from what has been observed typically in the metal rings
by many authors,\cite{WW,EFIELD} and also from the previously
reported results in GaAs/Al$_x$Ga$_{1-x}$As heterostructure
rings.\cite{Timp,ford88,PGNdV}.  There is a strong
correlation in the oscillation phase and amplitude throughout the
magnetic field range.
This implies that the scattering has been eliminated considerably in
these devices, or that the scattering event are ``gentler".  The latter
prospect is consistent with the disappearance of noise in filled
subbands.\cite{LBnoise,Timpqpc}
Different sweeps at the same gate voltages are in good agreement.
The reproducibility of two traces at a fixed $V_g$ is 99\% with a time delay of
5 hours between the up and down trace in magnetic field, which is comparable
to earlier work on GaAs/Al$_x$Ga$_{1-x}$As\cite{Timp}.
If we use the same data and subtract the smooth background (see below), the
envelope of
the $h/e$ oscillations, which account for $\sim$ 5-15\%
of the total conductance,
has a reproducibility of 70-90\%.
In the following we will concentrate on the data which is ordered and
reproducible.

The exact reason that the devices get into one or the other state is not
clear yet. A more systematic study of device processing and low temperature
transport is needed.
It is suggestive that under conditions where good plateaux exist in $G(V_g)$
the AB effect is cleaner.
For two of the samples used in this paper, the initial cooling (300K to
0.04K) obtained noisy data, one with very active time dependent conductance
fluctuations on the time scale of seconds, which
almost totally buried the AB effect
(nevertheless reasonably good conductance plateaux were seen at $T \sim 1K$
in the same cool-down).
After bringing the samples to $300K$ {\em in the dark} and recooling back
to $0.04K$, the {\em same} sample was in quiet, ordered state, and clean
AB oscillations were then observed.
Once the sample is in the ordered state, it
is fairly stable unless it is raised to a very high
temperature ($\gg 4.2K$) or suffers an electrical shock.
Based on the theory and experiments on universal
conductance fluctuations, this is not so surprising,
because only few active impurities are enough to kill the
correlations and even the AB effect itself,\cite{AS}
even when most of the conductance channel is ballistic.
The regions near the ports are especially critical and an inconvenient
impurity configuration there can have a huge effect,\cite{SZ} and this just the
region where the strongest electric fields appear during transport experiments.
So we speculate that above
randomness and changes from ordered to random ``states''
is related to the movement of a few
impurities or defects, probably near the ports.
With the view towards the ultimate goal of pure quantum waveguides,
these results imply that the task is even more formidable.
Not only a very clean channel is needed, but also the elimination
of all back-scattering within $\sim l_\phi$ of the ports.

Each trace (random or ordered)
can be described by a field dependent smooth background
resistance summed with Aharonov-Bohm oscillations of frequencies h/e,
h/2e, and so on.
The background resistance (the smooth line through the
oscillations of curve a in Figure 2) is calculated by averaging
the original data in every
$\Delta B = 0.005T$ interval of magnetic field.
This background resistance can be attributed to two parts.
One is the overall parabolic component, which we believe to be
the magnetic steering of electron away from head-on collisions
with the inner wall of the ring.  Other physics might contribute, however,
for instance, the electron-electron interactions can cause a similar
negative magnetoresistance\cite{paalanen83}.  The electron-electron
interaction should be strongly enhanced by the 1D nature of the transport,
but (perhaps) suppressed by the screening from the gate and complicated by
the donor charges.\cite{lutt}
However, the cause of the large negative magnetoresistance will not affect the
analysis to follow.
The second component of the smooth background resistance
comprises the broad peaks
at $\sim \pm 0.07T$, which agree well with a model of trapped
classical orbits in the loop.\cite{liu93}
The peaks in resistance arise when the cyclotron orbit of a
individual mode matches
the size of the ring, so that the electrons in that mode are
guided away from the outlet port and remain in the ring.
A peak appears for each individual value of forward
momentum, i.e. for each occupied transverse
mode, so at the same Fermi energy more than one trapping peak could be seen.
In this figure the trapping peaks, except for being dressed by the AB
oscillations, are not different from the equivalent data taken at high
temperatures.
The lack of temperature dependence provides further support for our
semiclassical trapping picture,
and rules out interference-related models for the resistance enhancement.

To be able to see the AB oscillations better, the smooth background
$G_0$ (reciprocal of the dark line) has been subtracted from the original
conductance in figure 2.
Since we will compare the AB effect from
a wide range of $V_g$ and $B$, in which
the background itself changes (typically) by a factor of $5$,
we will study the relative conductance change $\Delta G /G_0$
(which is also in line with the usual perturbative treatment of AB effect),
and this relative conductance oscillation for the top curve in
Figure 2 is shown in Figure 3a (Figure 2 is cut to $\pm 0.1T$ to
show the detailed structure of oscillations).
The envelope function here is mainly from the contribution of $h/e$ frequency.
The average spacing between two adjacent nodes is
$0.07T$, spanning about 34 fundamental AB oscillations.

It is straightforward to realize that even in the ideal case
when absolutely no random scattering is involved,
the envelope function will {\em not} be featureless; instead it will be
governed by some physics which was not emphasized in the diffusive case.
First, due to the finite width of the ring arms, different modes
with different {\em spatial}
distribution across the width of the wire will encircle different amounts
of flux and, therefore, have different AB frequencies.
A simple calculation of the flux difference between the first mode and second
mode in a square well leads to a result $\sim \pi (r_o - r_i)^2 B/ 4$,
where $r_o$ and $r_i$ stand for the outer and inner radius of the ring.
For our geometry we estimate a 4\% of difference between the frequencies, and
for higher modes our simple argument will give somewhat smaller result.
The result of the composition of two close frequencies is periodic beating
of the oscillation amplitude rather like that in Fig.\,3a.
If we attribute the change in amplitude to such a mechanism, then
typical h/e frequencies differ only by $\sim 3\%$, which is in reasonable
agreement with our estimate.  From this point of view, Fig.\,3a is very
close to what we can see from an ideal solid state interferemeter.
We note, however, that even when only one mode is populated there is still
some beating of the interference amplitude, although over a longer
field scale.  Another more plausible explanation is directly related to
the cyclotron orbit trapping.  When the electron is guided
away from the outlet, the amplitude of the oscillations, which result from
interference of trajectories escaping the ring, will be suppressed too.
As a result, a node will develop in the AB oscillation amplitude
accompanying a trapping peak in the resistance.
This naive model is not completely supported by
curve a in Figure 2a where the envelope nodes do not always line up with the
trapping peaks, but the average period is never-the-less on the right ballpark.

Ordered data like curve a in Figure 2 occur at specific values of $V_g$,
while at other $V_g$ the data are not as satisfying, but
as long as the sample stays in the quiet state,
the typical data are much more ordered than the random case (curve b) in
Figure 2.  So we generally characterize our ``ordered"
data as in an intermediate scattering regime, where analysis of
the envelope functions based purely on ballistic transport is not entirely
adequate.  But, as we will see, statistical methods are appropriate
and useful in this special regime.

\vspace{10pt}
\begin{center}
{\bf Comparison with mode counting steps }
\end{center}
\vspace{10pt}

Before using more sophisticated methods, we will directly look at the Fourier
transform (FT).  Figure 3b shows the FT amplitude of the data in Figure 2a.
Since the smooth background has been subtracted, the
zero frequency component in FT spectrum is not present.
The effective cut-off
``frequency'' ($1/\Delta B$) due to this filtering is $\sim 200 T^{-1}$.
To compare the FT spectrum with the the mode counting steps, systematic
measurements has been conducted for the double ring sample.

In Fig.4 the zero-field conductance $G$ versus $V_g$ at the temperatures
$2K$, $0.74K$ and $60mK$ is shown.
The conductance staircase is clearest at $0.74K$, and
at $60mK$ an oscillatory feature dresses the staircase.
The oscillations in $G(V_g)$ always accompany the AB effects.  At temperatures
where the staircase was smooth ($T > 1K$), no (or very small) AB
oscillations were observed.
In this figure, we also show the $60mK$ data after subtracting the
$0.74K$ sweep.  The average magnitude of the $G(V_g)$ oscillations agrees
with the magnitude of the AB effects at a given temperature.  This leads us
to relate the $G(V_g)$ oscillations to quantum interference.\cite{kltheory}
These features differ from the AB effects in that they are not
caused by the magnetic flux, they are related instead
to the change of the Fermi wavelength with changing electron density.
{}From the AB effects, we know that the dominant contribution to
the interference oscillations is from the fundamental $h/e$ signal.  This
frequency results from partial waves from the two ring arms interfering
at the outlet. If there is a difference in the two arm lengths, when
the Fermi vector changes, the phases accumulated on the trajectories through
each arm will be
different, the difference being $\Delta l \,\Delta
k_F $.\cite{kltheory,CJBFk}
The average period of the coherence oscillations in $G(V_g)$ is
$\Delta V_g = 3mV$.
Using our previous Shubnikov-de Haas measurements of
Fermi energy in similar samples,\cite{liu93} we have estimated that the
corresponding change in Fermi wavelength is $\Delta k_F=40 \mu m^{-1}$.
{}From $\Delta l \Delta k_F = 2\pi $, we have $\Delta l /l = 2\pi /
( \Delta k_F \pi r ) \sim 6\%$, or $\Delta l = 0.15 \mu m$.
This is certainly plausible.  The existence of so many oscillations (about
30 total), however,
implies that the simple picture above must be modified, since it would account
for about 10 oscillations at most.  If the modes are distinct within the wire
then each subband can contribute separate oscillations, which is consistent
with the increase in the number of oscillations per plateau as more subbands
are populated.  Another possibility is that of Fabry-Perot interference between
the inlet and outlet ports, which has been reported in similar
structures\cite{smithFP}.  Further experiments are required to sort out
the details as well as the rest of the phenomenology.

In Figure 5a and 5b we show the Fourier spectrum of
the relative AB oscillations at a series of
gate voltages.
The surface represents 32 magnetoconductance measurements equally spaced
between $V_g=0.5975V$ to $0.6700V$ and interpolation between successive curves.
Three peaks ($h/e$, $h/2e$ and $h/3e$) are clearly seen throughout
the range of
gate voltage, and the $h/4e$ peak can be seen at high $V_g$.
The average frequencies for the first three peaks are
$506.6 \pm 6.4 T^{-1}$, $1011 \pm 11T^{-1}$ and $1511 \pm 18T^{-1}$
respectively, which scale fairly linearly with the harmonic order as
expected.  To obtain the frequencies, each harmonic peak at a
specific $V_g$ is fitted with a Gaussian. There is {\em no}
systematic dependence of the frequencies on $V_g$ for any of the peaks.
These frequencies are typically larger than corresponding frequencies
from the single ring,
where $h/e$ oscillation is at $f= 482.2 T^{-1}$. We speculate this is due to
the widening of the ring at the ``hip" region.

In the same figure we include a panel at the left containing $G(V_g)$ at 0.06K
and for comparison, the (negative of the)
transconductance $-g_m = -d G/  d V_g$ has also been
plotted in the same frame.  In $-g_m$ the plateaux of $G$
correspond to plateaux between -50 to 50, and the risers
appear as valleys.  In this range of $V_g$ the
conductance $G(V_g)$ increases from $\sim 0.94 e^2/h$ to $\sim 4.2 e^2/h$,
corresponding to switching-on of modes 1 to 4.
We can see that, despite the large change in $G(V_g)$, the average
relative AB oscillation $\Delta G/G$
for $h/e$ is approximately constant (although there are order of magnitude
fluctuations)
implying that the contributions to $h/e$ frequency from each mode are
about the same.
In contrast, there is an slight increase in $\Delta G/G$ with $V_g$ for the
higher harmonics, especially $h/4e$.
In principle, this is not expected.
At low $V_g$ our resolution in $\Delta G$ decreased
because the detected current was smaller, so it is possible that the
apparent increase in the fourth peak is an instrumental artifact.
Other possible explanations include a longer phase coherence
at higher $V_g$ or an effect of the intermode scattering when many modes
are populated.

For the $h/e$ frequency, there is an obvious correlation between
the spectral density and the population of the modes.
When the $B=0$ conductance is at the center of a conductance plateau,
the AB oscillations are stronger, and on the risers weaker.

Now examine the scattering processes when a new mode is turned on.
Because the subband bottom mainly comprises small $\em k$ states,
it is sensitive to imperfections of the conducting channel.
As a result, the scattering is stronger when a mode is newly populated, which
is the reason for the finite width of the risers in $V_g$.
According to the B\"{u}ttiker-Landauer formula,
small angle intra-band scattering has very little effect on the conductance.
So we can think the gradual risers are merely
manifestations of heavy inter-band or large-angle intra-band scattering
(generally are called back-scattering).
Both of these scattering mechanisms
will reduce the $h/e$ AB oscillation size (their roles in general, and for
$h/2e$ will be discussed later).
Following this line of thinking, a good staircase in
$G(V_g, B=0)$ (not counting the fine interference pattern in curve d in
Figure 4) foretells observation of a periodic oscillation of the scattering
strength, and an observation of a correlation between $G(V_g)$ and
the AB magnitude is thus {\em inevitable}.
The lack of observation of such correlations in earlier experiments in
heterojunctions is consistent with the absence of a clear zero B
staircase in their
experiments.\cite{Timp,GAloops,ford88,CJBFk,TimpQHE,PGNdV}
The above analysis is based on a static scattering potential, and not
time dependent events which may serve as a phase-breaking source, but not
a conductance killer (eg. spin-spin interactions).  In this case
we may see a staircase at $T \sim 1K$, but at lower temperature might not
be able to observe any correlations in the AB effect (or in fact
any phase coherent response at all).  This may be
the reason for poor AB oscillations in the ``random'' state (see above).

The discussion in the last paragraph
implies that the quality of correlation closely depends
on how clean the staircase is.
We note that the low temperature oscillations at $B=0$ (curve d
in Figure 4) contribute a significant amount to the conductance
($\sim 10\%$).  The non-ideal correlation may be
attributed the substructures in the curves of Figure 4.
One should in principle self-consistently model the AB interference
patterns with
the dressed $B=0$ transmission coefficient.\cite{kltheory}

As a final remark we note that,
if, starting in Figure 5 at a peak in the
FT at $h/e$ frequency, we move towards
either side of $V_g$, the peak height decreases smoothly while
the peak frequency drifts.  As a result, rather than uncorrelated
random spikes, hills of significant footprint are seen. The gate voltage
correlation range among the hills
the hills is $\sim 20mV$, i.e. a range in which almost a full mode
is turned on.  This correlation with
gate voltage correlation may be a manifestation
of the response of a particular mode evolving its detailed charge distribution
as $V_g$ change, but more detailed investigation of this physics is required.
Another consequence of the frequency shift is that the peaks
move as $V_g$ changes, so if we show a cross-section
at any particular $f$, the global correlation seen in Figure 5 will
{\em not} be apparent.
For the higher harmonics ($h/2e$, $h/3e$ and $h/4e$), no obvious
correlations can be seen between the peak heights and the plateaux in
$G(V_{g},B=0)$.

\vspace{10pt}
\begin{center}
{\bf Analysis of the envelope functions}
\end{center}
\vspace{10pt}

Besides studying the magnitude of the AB oscillations, we also investigated
the patterns of the envelope functions with respect to the subband
populations.
Similar data has been discussed previously
for the disordered (metallic) limit.\cite{WW}
One expects the interference contribution to the
conductance to be of the form\cite{kltheory,ImryStone}
\begin{equation}
\Delta G = \sum _{j=0} ^{\infty} G_j (B,V_g )
cos \left[ \frac{2\pi j \Phi}{(h/e)} + \alpha _j (B,V_g) \right],
\end{equation}
where $\Phi$ is the average amount of magnetic flux enclosed by the
electron trajectories encircling the ring,
while $G_j$ and $\alpha _j$ are the envelope function and phase that account
for individual harmonics.  Information about the
details of the scattering are contained in $G_j$ and $\alpha_j$ while
the oscillatory components contain only the size of the ring.
For metal samples
these functions have been random in B.
The correlation scale for
a envelope function $G_j$ is $B_c$, which in the diffusive case, is just the
field scale to introduce a flux h/e into the sample (i.e. the ports and the
arms of the ring).
But this diffusive description for $B_c$ do not apply to our data.
Because even though the effective width of the wire does increase with $V_g$,
but it is too small, only 40\% \cite{liu93}, to account for what
we have observed in our samples; in addition for our data the $B_c$
changes periodically rather than monotonically (see below).
Here we will borrow this vocabulary, such as envelope function $G_j$ and
correlation field $B_c$,
to describe our data.  And for the ensuing analysis of the data we will
explore the degree of ``order" in the envelope functions $G_j$, through
their autocorrelation functions.

Now consider the two contrasting curves in Figure 2 first.
In order to analyze the envelope $G_1$ for the $h/e$ oscillations, we
use a Gaussian filter and reverse Fourier transform each $h/e$ peak back
into B space. Figure 6a illustrates the result for $h/e$ oscillations
from spectrum in Figure 3b.  The $h/e$ signal accounts for about 80\%
of the total oscillation magnitude in Figure 3a.
To ensure the information in the chosen
peaks are fully included, generous filters of half-width
$> 100T^{-1}$ around the center frequencies were used.  These
are at least 3 times bigger
than the widths ($15 T^{-1}$ and $30T^{-1}$) of the h/e and h/2e peaks
calculated by fitting Gaussians to the peaks.
To reassure the validity of the filter, we have done a FT for Figure 6a.
If we plot the result in the same picture with 3b, no difference could be
seen between these two curves in the region $\pm 100 T^{-1}$ around the
center frequency.
The same filter width was used throughout the analysis.
The same operation for the bottom curve in Figure 2 obtains
the result in Figure 6b.
Both curves contain AB oscillations of the same average frequency.
The differences between the two curves are in the envelope function $G_1$
(dark curves that bound the oscillations),
in both the amplitude (a factor of $\sim 5$ even in the {\em relative}
conductance) and in the correlation scales.

In Figure 7 we have calculated the autocorrelation functions (solid lines)
for $G_1$ in Figure 6.
The huge difference between the two envelopes is reflected here.
For envelope 6b, the oscillation amplitudes at a particular value of B
are not correlated with those at other places on the B axis, so its
autocorrelation function for the envelope (7b) is mainly Gaussian-like:
a single monotonically decaying peak signifying random correlations.
The long-dashed line is the autocorrelation function calculated from the
average value of the envelope, which is consistent with the constant
offset.
In contrast for $V_g = 0.15V$ a much more regular envelope function seems
to prevail. The autocorrelation function for this envelope contains a
decaying peak and {\em a regular} oscillatory ``tail",
which indicates a non-random pattern in the envelope function.

In general, the shape of the correlation function can be understood as
follows.
Because the envelope function basically consists
of the summation of two parts,
$G_1(B) = P(B) + Q(B)$,
where $P(B)$ is a periodic function, and $Q(B)$ a random function.
The autocorrelation function of $G_1(B)$ can then be calculated as,
\begin{eqnarray}
C(G_1) & = & \int [P(B) + Q(B)][P(B+ \Delta B) + Q(B+ \Delta B)] dB
\nonumber \\ & = & C(P) + C(Q) \\ && + \int
\nonumber [Q(B)P(B+ \Delta B) + P(B)Q(B+ \Delta B)] dB \\
& = & a \cos (2 \pi B/B_{c1}) + c \exp (-B^2/B_{c2}^2) + constant
\nonumber
\end{eqnarray}
In the derivation, we have assumed that $P(B)$ and $Q(B)$ are
uncorrelated, so the cross-term only contribute a insignificant constant.
The autocorrelation for $P(B)$ is simply
a $cos$ function, $C(P)= a \cos (2 \pi B/B_{c1})$;  and
for the random function $Q(B)$ a Gaussian $C(Q)= c \exp {(-B^2/B_{c2}^2)}$.
The final constant is from all three terms (cross-term, $C(P)$ and $C(Q)$),
since both $P(B)$ and $Q(B$ has nonzero average values.
$B_{c1}$ is the ``beating'' period, and $B_{c2}$ the random correlation scale.
We acknowledge that there are some problems associated with performing the
calculation in a finite field range.
The consequence is that none of the above three terms would be ideal, the
random function $Q(B)$ will produce oscillatory features in both the
 cross-term
and in Gaussian term. This is clearly seen in Figure 7b; there are oscillatory
features in the ``tail'', which would disappear if the field range were
long enough.  But we also realize that if the correlation $B_{c2}$
does not exceed the long range correlation field $B_{c1}$ (which is what always
assumed for the experimental data anyway), only $C(P)$ gives the long-range
correlation field $B_{c1}$, which is what we are primarily interested in the
analysis.

If we use eq. (2) to fit the two correlation functions in
Figure 7 (dash lines), it yields $B_{c1} = 0.069 T$, $B_{c2} = 0.015 T$
for curve a, and $B_{c1} = 0.012 T$, $B_{c2} = 0.0024 T$ for curve b.
$C(Q)$ is restricted to small $\Delta B$; on the
other hand, $C(P)$ persists to large $\Delta B$.
Here beating period $B_{c1}$ represents the long range order of
the envelope function.
For both curves we can see the fitted $B_{c1}$ are consistent with the
periods we get from the original envelope functions in Figure 6.
So the beating period $B_{c1}$ can be used as a parameter to
characterize the ``order'' of the envelope function quantitatively.
Even for the metallic case 7b, which is outside the realm of
validity for formula (2), the fit yields correctly, at least qualitatively,
a very small value of $B_{c1}$, but
the fit quality is rather poor, indicating that
there is no long range order in this correlation function.
In the very ordered case such as 7a, there might be an even longer $B_{c1}$
if the measurement were extended
to wider range of field, because the window size $\pm 0.2 T$
puts a upper limit of $0.2T$ on $B_{c1}$.
Unfortunately, this cannot be done because
depopulation of the subbands and Shubnikov-de Haas oscillations and finally
the quantized Hall effect set in, and they cause dramatic changes in the
physics behind the AB oscillations, which
goes beyond the physics that this paper set out to to study.

We have performed the same calculation for
all three samples.  We notice that for coupled rings (2 rings and 4
rings), equation (2) does not always fit the data, and the discrepancy worsens
as more modes are occupied.  Representative data from
2-ring sample at $V_g=0.660V$ data shown in Figure 8.  The autocorrelation
function (solid line) contains two oscillation periods.
Obviously a single beat frequency is not
adequate in these cases.  Instead of invoking a more
complicated model, we will keep using (2) to fit the original data by seeking
a local minimum in the sum of squares of differences starting from
different initial conditions.   This method handily yields
the two beat frequencies as shown (dotted and dashed lines) in
Figure 8, where the two periods are $0.081T$
and $0.019T$, respectively.  The average beating periods $B_{c1}$ for $G_1$ of
the three samples are collected in Table 1.  In cases where
two frequencies exist, they were considered with equal weight in the
statistics.
\vspace{1cm}

\hspace{2.5cm}
\begin{tabular}{||c|c|c|c||} \hline \hline
\hspace{2mm} SAMPLE  \hspace{2mm}&\hspace{5mm}
$\overline{B_{c1}}\hspace{5mm}$ &\hspace{2mm}
\# of $V_g$\hspace{3mm} &\hspace{2mm}mode range \hspace{2mm}\\ \hline
1  ring & 0.078T&8&1 $\rightarrow$ 2\\ \hline
2 rings & 0.045T&15&1 $\rightarrow$ 2\\ \cline{2-4}
 & 0.038T&32&1 $\rightarrow$ 4\\ \hline
4 rings & 0.037T&8&1 $\rightarrow$ 2\\ \hline \hline
\end{tabular}

\vspace{1cm}

{}From this table, we can see that, in the same mode population range, the
average beat period tends to decrease as the number of rings increases.
For the 2-ring sample, the only sample
for which a relatively wide mode population
range was measured, the average beat period tends to decrease as more
modes are populated.
Among all the samples, only the data from 2-ring sample allows us to make
a detailed comparison between the AB effect and $V_g$.
The beating periods for $h/e$ oscillations are summarized in Figure 9b.
In the calculation the fitting results are constantly checked ``by eye''
against the original envelopes to make sure that the
fit results are reasonable.
The biggest and the smallest period differ by almost an order of magnitude.
In cases where two periods exist, the extra one is plotted as a `x'.
The dashed line connects all points (in cases where there are two
points, the average is used) together, and the
solid line is a smoothed rendition
of the dashed line
Again $G(V_g)$ and $g_m$ are shown in Figure 9a.
On the plateaux, the correlation functions have
relatively long periods $B_{c1}$, and shorter periods on the risers,
which coincides with the relationship between FT peak amplitude and $G(V_g)$
in Figure 5.

We have also performed the reverse Fourier transform for $h/2e$ oscillations,
and the resulting $B_{c1}$ is shown in Figure 9c.
However, the overall trend seems to be about
the opposite of that for $h/e$
(There is no similar correlation for the FT amplitude as a function of $V_g$
in Figure 5.).
This result came a little surprising to us at beginning, we do not rule out
the possibility of an artifact.
But it is also plausible that the trend is an indication of real transport
physics.
We have discussed the role of back-scattering in quenching
the $h/e$ oscillations.
Now consider the case of large-angle, intra-band scattering,
if a $\em k$ is reflected into the $\em -k$
state in the same subband, the phase
coherence is retained.  This kind of scattering kills the $h/e$
oscillation if the scattering takes place either
inside the ring or in the ports.
If the scattering occurs in the ports it could
{\em enhance} the $h/2e$ oscillation amplitude, but
the overall effect of back-scattering on $h/2e$ is, at least, not clear.
The simple characterization of scattering
as ``elastic" and ``inelastic'' (the latter usually synonymous with
``phase-breaking''), which
so far has guided our thought on diffusive transport, is no longer sufficient.
Instead more detailed study of the roles of specific back-scattering
mechanisms on different AB oscillation components is
necessary.

\vspace{10pt}
\begin{center}
{\bf Summary}
\end{center}
\vspace{10pt}

We have performed a careful experiment on GaAs/Al$_{x}$Ga$_{1-x}$As rings
to study the correlation between the $B = 0$ subband population
and the AB oscillation amplitude at low magnetic fields.
Strong correlations are observed between the one-dimensional
subband populations and the
Fourier amplitude of the oscillations.
There is also a correlation with the degree of order in the
envelope function as judged through its autocorrelation function.
These samples have shown
improvements on eliminating most of the impurity scattering, pointing towards
the possibility of purely ballistic solid state interferometers.
Questions for further theoretical and experimental investigation include
whether or not
the modes contribute independently to the interference patterns and
how classical mechanisms (such as scattering in the port junctions
and orbit trapping) affect the
envelope function.

\vspace{10pt}
ACKNOWLEDGMENT: This work was supported
by IBM and the Microelectronics Center of
North Carolina. We thank K. Li, V. Long, and Y. Wang for
their assistance during the
measurements and D. Kern and S. Rishton for help with lithography.

\newpage

\figure{
Schematic drawing with lithographic dimensions for a two-ring sample.
The loops and ports are to scale, but the large area regions are not.
The region comprising the rings and the ports
is covered by a Ti/Au gate.
}
\figure{
The magnetoresistance for the single ring at the gate voltage of
$V_g = 0.15V$ (a) and $V_g = 0.3V$ (b).
The smooth line through the $0.15V$ data
is the background resistance
(its reciprocal is $G_0$ which will be used later), calculated
by averaging the original data in $0.005T$ intervals. The peaks near
$\pm 0.07T$ are due to trapping of cyclotron orbits.
}
\figure{
(a) $\Delta G/G_0$ is the relative conductance oscillations calculated
from the subtracting curve a in Figure 2 from the averaged (heavy) line.
(b) The Fourier transform amplitude
of $\Delta G /G_0$ from (a).
}
\figure{
$G(V_g)$ at several temperatures and $B=0$ recorded upon a second
cool-down with a considerably shifted threshold voltage.  The
temperatures for the curves are:
a: $T = 2K$, b: $T = 0.74K$, and c: $T = 0.06K$.
The difference between c and b, which is attributed to
interference effects from changing the Fermi vector $k_F$, is plotted as d.
}
\figure{
The Fourier transform of $\Delta G/G_0$ as a function of gate voltage for
the double-ring sample, in frequency ranges
from $300T^{-1}$ to $1300T^{-1}$ (a), and $1300T^{-1}$ to $2300T^{-1}$ (b).
The vertical axis in (b) is the same as for (a).
The surfaces are constructed from 32 equally-spaced
measurements between $V_g=0.59V$ and
$V_g=0.67V$ recorded on the second cool-down.
Four peaks are seen at the frequencies $506.6 \pm 6.4 T^{-1}$,
$1011 \pm 11T^{-1}$, $1511 \pm 18T^{-1}$ and $\sim 2020 T^{-1}$.
The $G(V_g,B=0)$ and (the negative of) its derivative
$-dG/dV_g$ versus $V_g$ are drawn on the z-y panel.
The correlation between the mode population and the height of the $h/e$
peak is
obvious; there are four main peaks in $h/e$ and the same number ``hills''
in $-dG/dV_g$.  The strongest $h/e$
oscillations are found around the conductance plateaux, i.e.
the zeros (``hills'') of $-dG/dV_g$.
}
\figure{
Inverse Fourier transforms of the $h/e$ peaks are calculated
to obtain the oscillation patterns for the upper
(a) and lower (b) curves in Figure 2.
The two patterns are defined by different amplitude
envelope functions $G_1/G_0(B)$, which are drawn as the smooth
dark lines along the oscillation maxima.
}
\figure{
The measured autocorrelation function (solid line) of the envelope functions
$G_1/G_0$ for the single ring for (a) the upper curve
and (b) the lower curve in Figure 6.
The dashed line in (b) is the autocorrelation function calculated
from the average value of the envelope, which is consistent with the
constant background offset.
They are fitted with eq. (2) (dotted line)
to obtain the beat periods $B_{c1}$ and decay scales $B_{c2}$.
}
\figure{
The autocorrelation of $G_1/G_0$ for the double ring (solid
line) at $V_g=0.660V$ (second cool-down).
It contains more than one beat frequency.  It was fitted with formula (2)
with different initial conditions to obtain the
periods $0.081T$ (dotted line) and $0.019T$ (dashed line).
}
\figure{
Comparison between the mode counting staircase $G(V_g)$ (solid
line) and transconductance (dotted line) recorded at $B=0$ (a)
and the beat periods for the $h/e$ (b) and $h/2e$ (c) peaks from
all of the measurements from Figure 5.
In (b) when two periods are found for a single correlation
function as in Figure 4, the extra one is shown as `x'.  The dashed lines
connect the beat periods (the average in cases there are two).  The solid line
is a smooth (Gaussian filtered) version of the dashed line.
In (b) there is a correlation to the staircase similar to that seen in
figure 5 for the average amplitude of the oscillations.
In (c) the opposite trend to (b) is seen.
}

\end{document}